\documentclass[twocolumn,floats,prb,aps,showpacs]{revtex4}
\usepackage[final]{graphicx}
\usepackage[dvipsnames]{color}
\usepackage{dcolumn}
\usepackage{hhline}
\DeclareGraphicsExtensions{.pdf,.eps}

\begin{document}

\title{Resonant hot charge-transfer excitations in fullerene-porphyrin complexes:
a many-body Bethe-Salpeter study}

\author{Ivan Duchemin}
\affiliation{INAC, SP2M/L$\_$Sim, CEA/UJF Cedex 09, 38054 Grenoble, France.}

\author{Xavier Blase}
\affiliation{Institut N\'{e}el, CNRS and Universit\'{e} Joseph Fourier,
B.P. 166, 38042 Grenoble Cedex 09, France. }

\date{\today}

\begin{abstract}
We study the neutral singlet excitations of the zinc-tetraphenylporphyrin and $C_{70}$-fullerene 
donor-acceptor complex within the many-body Green's function $GW$ and Bethe-Salpeter approaches.
The lowest transition is a charge-transfer excitation between the donor and the acceptor
with an energy in excellent agreement with recent constrained density functional theory 
calculations. Beyond the lowest charge-transfer state, which can be determined with simple 
electrostatic models that we validate, the Bethe-Salpeter approach provides the full excitation
spectrum. We evidence the existence of hot electron-hole states which are resonant in 
energy with the lowest donor intramolecular excitation and show an hybrid intramolecular and 
charge-transfer character, favoring the transition towards charge separation. Such findings,
and the ability to describe accurately both low-lying and excited charge-transfer states, are
important steps in the process of discriminating ``cold" versus ``hot" exciton dissociation 
processes.
\end{abstract}


\maketitle


\section{Introduction}

Photo-induced charge-transfer excitations are at the heart of a large variety of physical
phenomena and related applications, from photovoltaic cells for renewable energy applications,
\cite{CTCrefs} to light-harvesting biological systems involved e.g. in the photosynthesis 
processes occurring in plants or bacteria. \cite{photosynthetic} In the former case of the 
so-called third generation solar cells such as all-organic devices or dye-sensitized Gr\"{a}tzel 
cells (DSSCs), the large binding energy of the photoexcited electron-hole pair requires the 
presence of a donor-acceptor interface in order to dissociate the bound excitons and produce 
the free carriers at the origin of the photocurrent. \cite{CTCrefs} Even though the mechanisms 
explaining such a charge separation are still very debated,
\cite{Muntwiler08,Ohkita08,Clarke08,Pensack09,Lee10,Bakulin12,Caruso12,Murthy12,Yost13} 
it is believed that 
important intermediate states are the so-called charge-transfer excitations, with the hole and
electron located respectively on adjacent donor and acceptor units (organic molecules, inorganic 
surface or cluster, etc.)  Understanding the structural and electronic factors that control the
transition from the photoinduced intramolecular (Frenkel) excitons, generated by light onto the 
donor molecules, to a bound charge-transfer (CT) state and subsequently to an unbound electron-hole
charge-separated (CS) state, is an important goal on the road of improving the quantum efficiency 
of solar cells.

From a theoretical point of view, the correct description of non-local excitations, such as
charge-transfer excited states with electrons and holes spatially separated, 
stands as a challenging issue. In particular, it is now well established that ``standard"
time-dependent density functional theory (TDDFT) approaches \cite{tddft} based on (semi)local 
exchange-correlation kernels, in the absence of exact exchange contribution,
fail in providing a correct description of such charge-transfer (CT) excitations. 
\cite{Dreuw03,Dreuw04} This has been demonstrated in the case of inter- and 
intra-molecular CT excitations, but also in the case of extended bulk Wannier excitons in 
semiconductors for which holes and electrons are separated by a large effective bohr 
radius. \cite{Botti04} Solutions involving a contribution from the exact Fock operator,
based on e.g. range-separated hybrids, \cite{Savin,Kronik12} are being actively developed 
and tested, with various choices for the range and percentage of exact exchange. 

In a recent study, \cite{Ghosh10} the lowest charge-transfer excitation in the zinc 
tetraphenylporphyrin (ZnTPP) donor associated with the $C_{70}$  fullerene acceptor
was studied using a density functional theory (DFT) approach within an elegant and
computationally efficient constrained formalism with a penalty functional forcing the 
jump of a chosen fraction of an electron from the ZnTPP highest occupied molecular orbital 
(HOMO) to the $C_{70}$ lowest unoccupied molecular orbital (LUMO).  It was shown in particular
that the related excitation energy of $\sim$ 1.8 eV in case of a full electron transfer was 
much larger than the $\sim$ 0.9-1.0 
eV energy obtained with TDDFT based on the PBE kernel. \cite{PBE} Even though the available 
experiment \cite{Mukherjee09} was performed in a solvent with a different porphyrin derivative,
the constrained-DFT result is in much better agreement with the range of experimental values.

Computationally very efficient and accurate, such DFT-based techniques \cite{Voohris06} 
cannot provide however the detailed excitonic (or absorption) spectrum at the donor-acceptor 
interface.  In particular, the energy position of the donor intramolecular (Frenkel) low-lying 
exciton  with respect to the charge-transfer excitonic spectrum, including the higher energy 
(hot) charge-transfer states, is a crucial information needed to understand the dissociation 
of the excitons generated on the donor side and that have diffused to the donor-acceptor interface. 
Mechanisms where intramolecular excitons evolve into charge-separated states through hot (high 
energy) CT states have been recently proposed  on the basis of experimental and theoretical data
\cite{Muntwiler08,Ohkita08,Clarke08,Pensack09,Bakulin12,Murthy12} so that the knowledge 
of the entire excitation spectrum (at least in a few electronvolts energy window) should
prove of crucial importance in a attempt to shed some light onto the mechanisms that may
explain the observed conflicting evidences on the role of cold or hot charge transfer excitations. 
Besides the energy spectrum alone, the related knowledge of the excitonic wavefunctions is a crucial 
ingredient governing e.g. the calculation of charge-transfer rates. Furthermore, in the case 
of the constrained DFT study of the $C_{70}$-ZnTPP complex, the underlying assumption of a 
full electron transfer may deserve closer inspection.

Recently, \cite{Lastra11,Blase11b,Baumeier12} charge-transfer excitations in small gas 
phase donor-acceptor complexes, associating tetracyanoethylene with several acene
derivatives, were studied using the many-body perturbation theory framework 
within the so-called $GW$ \cite{Hedin65,Strinati80,Hybertsen86,Godby88,Onida02,Aulbur}
and Bethe-Salpeter (BSE) \cite{Sham66,Strinati82,Rohlfing98,Shirley98,Albrecht98} 
formalisms. It was shown in particular \cite{Blase11b,Baumeier12} that an excellent 
agreement with experiment \cite{Hanazaki72} could be obtained for the lowest charge-transfer 
excitation, with a mean absolute error of 0.1 eV. This accuracy matches that of
the latest range-separated functional TDDFT studies with the range parameter adjusted 
to provide an accurate quasiparticle energy gap.  \cite{Stein09} 
The same formalism was used to study intramolecular charge-transfer 
excitations in a family of molecular dyes for DSSCs, the coumarins, \cite{Faber12}
showing a mean absolute error of the order of 0.06 eV as compared to quantum 
chemistry coupled-cluster \cite{Kurashige07} or tuned range-separated hybrid TDDFT 
calculations. \cite{Wong08,Stein09,Rocca10} 

In the present work, we report the detailed excitonic spectrum of the $C_{70}$ fullerene 
associated with the zinc tetraphenylporphyrin (ZnTPP). This study is performed within the
many-body Green's function $GW$ and Bethe Salpeter formalisms using the same cofacial geometry 
as that used in the constrained DFT study of Ref.~\onlinecite{Ghosh10}.
Our results are in excellent agreement with constrained DFT calculations for the lowest 
lying $CT_0$ charge-transfer excitation, but provide further the entire excitation spectrum.
It is shown that several hot electron-hole excited states lie in between 
the lowest intramolecular ZnTPP$^*$ donor excitation and the CT$_0$ exciplex.
Of peculiar interest, several hot states with hybrid 
intramolecular and charge-transfer character are found to be resonant in energy with 
the photo-induced ZnTPP$^*$  exciton, an important finding in the perspective of the
recently proposed scenarios where charge-separation may occur directly through delocalized 
hot charge-transfer states.  

\section{Methodology}

Originally developed and applied at the \textit{ab initio} level in the mid-sixties to 
extended solids, \cite{Aulbur} 
the many-body Green's function $GW$ approach, aiming at providing accurate single-particle 
quasiparticle energies, is now being extensively tested for molecular systems in order
to assess its accuracy and limitations.  
We briefly sketch here the basic equations, starting with the eigenvalue equation:

\begin{eqnarray*}
 \left( { -\nabla^2 \over 2 } + V^{ion}({\bf r}) + V_H({\bf r}) \right) 
       \phi_n^{QP}({\bf r}) &+&  \\
      \int \Sigma({\bf r},{\bf r}';E_n^{QP}) \phi_n^{QP}({\bf r}') d{\bf r}' 
    &=& E_n^{QP}  \phi_n^{QP}({\bf r}), 
\end{eqnarray*}

\noindent where $V^{ion}$ and $ V_H$ represent the ionic (pseudo)potential and Hartree potential. 
As a central ingredient, the self-energy $\Sigma({\bf r},{\bf r}';E_n^{QP})$, which accounts for
exchange and correlation, is non-local and energy-dependent.
In the $GW$ approximation, the self-energy reads:

$$
 \Sigma^{GW}({\bf r},{\bf r}';E) = { i \over 2\pi} \int d\omega e^{i \omega 0^+}
   G({\bf r},{\bf r}';E-\omega) W({\bf r},{\bf r}';\omega),
$$ 

\noindent 
where $G$ and $W$ are the one-particle time-ordered Green's function and the screened Coulomb potential:

\begin{eqnarray*}
   G({\bf r},{\bf r}';\omega) &=& \sum_n { \phi_n({\bf r}) \phi_n^*({\bf r}')  
\over \omega - E_n + i0^{+} sgn( E_n - E_F ) }    \\
   W({\bf r},{\bf r}';\omega) &=& \int d{\bf r}' {\varepsilon}^{-1}({\bf r},{\bf r}'; \omega)
   V^C({\bf r},{\bf r}').
\end{eqnarray*}

\noindent
$V^C$ is the bare Coulomb potential and $E_F$ the Fermi level.
The $(\varepsilon_n,\phi_n)$ eigenstates are typically density functional theory Kohn-Sham
eigenstates. The inverse dielectric matrix $\epsilon^{-1}$ is calculated within the
random-phase approximation.

Our calculations are performed with the {\sc{Fiesta}} code,
\cite{Blase11b,Blase11a,Faber11a,Faber11b,Duchemin12} 
a recently developed $GW$-$BSE$ package based on a Gaussian auxiliary basis expansion of the 
non-local two-body operators such as the dielectric susceptibility, the bare or screened 
Coulomb potentials, combined with resolution of the identity techniques.  \cite{Ren12}
Dynamical correlations are calculated exactly using a contour deformation approach
\cite{Godby88,Farid99} without any plasmon pole approximation, performing the frequency
integration along the imaginary axis, with additional contributions from the poles of the
$G({\bf r},{\bf r}';E-\omega)$ Green's function. The correlation contribution 
then reads:

$$
\Sigma_c^{GW}({\bf r},{\bf r}';E) =  \sum_n  \phi_n({\bf r}) \phi_n^*({\bf r}') 
     {\nu}_n({\bf r},{\bf r}';E), 
$$

\noindent with, introducing $\tilde{W}=W-V^C$ and $\Theta$ the Heavyside function:

\begin{eqnarray*}
 {\nu}_n({\bf r},{\bf r}';E) &=& \tilde{W}({\bf r},{\bf r}';\varepsilon_n -E)
\left[ \Theta(E-\varepsilon_n) - \Theta(E_F-\varepsilon_n) \right] \\
  &-& \int_0^{+\infty} { d\omega \over \pi }  
      {E-\varepsilon_n \over  (E-\varepsilon_n)^2 + \omega^2} 
      {\tilde W}({\bf r},{\bf r}'; i\omega)
\end{eqnarray*}

The initial Kohn-Sham eigenstates are obtained from the  {\sc{Siesta}} 
package \cite{siesta} with a well-converged triple-zeta plus double polarization 
basis (TZ2P) \cite{basis1,basis2} and standard norm-conserving pseudopotentials
 \cite{pseudos} combined with the local density approximation (LDA) \cite{lda} for the 
exchange-correlation functional.  
Convergence tests can be found in the Appendix where we show that the differences between
excitation energies obtained with the TZ2P basis and a reduced TZP (single polarization) 
basis are smaller than 15 meV.
All empty states are included in the summation over 
transitions involved in the calculation of the independent-electron susceptibilities.  
As shown recently in the study of charge-transfer excitations in several $\pi$-conjugated 
organic donor-acceptor systems, \cite{Blase11b} the present 
scheme provides results in excellent agreement with planewave-based $GW$-$BSE$ studies performed 
with the Yambo code \cite{Yambo} exploiting Kohn-Sham eigenstates generated with the Quantum-Espresso 
package \cite{Espresso}. The auxiliary basis consists of six Gaussians per (\textit{s,p,d})-channel
with an even-tempered distribution of decay coefficients ranging from $\alpha_{min}$=0.1
to $\alpha_{max}$=3.2 bohr$^{-2}$.  Extensive convergence tests as a function of the auxiliary 
basis size can be found in a recent study of a photosynthetic donor-acceptor complex.
\cite{Duchemin12}

We go beyond the standard ``single-shot" $G_0W_0$ calculation by achieving 
self-consistency onto the quasiparticle eigenvalues, leaving the Kohn-Sham wavefunctions
unchanged. Such a scheme has been shown for molecular systems to lead to quasiparticle energy 
gaps, between the highest occupied (HOMO) and lowest unoccupied (LUMO) molecular orbitals, in
better agreement with experiment than the standard $G_0W_0$ calculation based on DFT-LDA or 
DFT-PBE eigenstates \cite{Blase11a,Sharifzadeh12,Abramson12,Hogan13,Hahn05} at a cost much reduced
as compared to full self-consistency. \cite{gwscf,hybridstart,Bruneval13} Similarly, $GW$-BSE CT 
excitation energies in reference donor-acceptor complexes were shown to agree much better with 
experiment if such a partial self-consistency is performed at the initial $GW$-level. \cite{Blase11b,Baumeier12}
The dependence of the final $GW$ quasiparticle energies on the starting Kohn-Sham eigenstates 
is further significantly reduced. \cite{reduction}  

While the $GW$ calculations provide accurate quasiparticle energies, including not only the 
correct HOMO-LUMO gap, but also a correct spacing and ordering within the occupied (unoccupied) 
states, the subsequent resolution of the Bethe-Salpeter equations  (BSE) aims at providing the 
correct neutral excitation energies, accounting in particular for electron-hole interaction. 
We  go beyond the Tamm-Dancoff approximation, including namely the coupling of resonant and 
non-resonant excitations. As shown in previous work, \cite{Ma09,Gruning09,Faber12} the 
Tamm-Dancoff approximation leads in the case of molecular (or nanosized) systems to an 
overestimation of the transition energies by as much as 0.3 eV.

The BSE Hamiltonian is most commonly expressed in the $|\phi_i^e\phi_j^h>$  product basis, 
where $(\phi_i^e)$ and $(\phi_j^h)$ are  unoccupied (electron) and occupied (hole) Kohn-Sham
states, respectively. In particular, the resonant contribution is composed of a 
``non-interacting" diagonal part ($H^{diag}$), a direct ($H^{direct}$) and an exchange 
contribution ($H^{exch}$) term, with:

\begin{eqnarray*}
 H^{diag}_{ij,kl} &=& \left( \varepsilon^{QP}_{i} - \varepsilon^{QP}_{j} \right) \delta_{ik} \delta_{jl} \\
 H^{direct}_{ij,kl} &=& - \int d{\bf r} d{\bf r}' \phi_i^e({\bf r}) \phi_j^h({\bf r}') W({\bf r},{\bf r}')
      \phi_k^e({\bf r}) \phi_l^h({\bf r}'), \\
 H^{exch}_{ij,kl} &=& 2 \int d{\bf r} d{\bf r}' \phi_i^e({\bf r}) \phi_j^h({\bf r}) v({\bf r},{\bf r}')
      \phi_k^e({\bf r}') \phi_l^h({\bf r}'),
\end{eqnarray*}

\noindent where the $(\varepsilon^{QP}_{i/j})$ are the quasiparticle ($GW$) energies,  while 
$W({\bf r},{\bf r}')$ and $v({\bf r},{\bf r}')$ are the (statically) screened and bare Coulomb
potential, respectively. It can be noticed in particular that the direct term does not cancel 
in the limit of spatially non-overlapping electron and hole states, an important remark concerning 
charge-transfer states. 

Due to the fast increase in size of the electron-hole two-body product $|\phi_i^e \phi_j^h>$ 
basis states, we restrict our BSE basis to transitions between the occupied states and 360 
unoccupied levels, that is namely empty states up to 20 eV above the HOMO level.  Convergence 
tests can be found again in Ref.~\onlinecite{Duchemin12} for donor-acceptor systems of similar size.  
We emphasize that the low-lying charge-transfer excitations, with much weight on the lowest occupied
and unoccupied Kohn-Sham electron-hole product states, converge extremely quickly with the size of 
the BSE basis, but this is not necessarily the case for intramolecular excitations or higher (hot) 
CT excitations. 

With the present scheme, charge-transfer excitations were shown to come within a mean absolute error 
(MAE) of 0.1 eV as compared to gas phase experimental data for a family of small donor-acceptor TCNQ-acene 
complexes, \cite{Blase11b} and within 0.06 eV  (MAE) as compared to coupled cluster (CC2) values for 
a family of coumarin dyes showing intramolecular charge transfer excitations. \cite{Faber12} Further, 
as emphasized here below and in Refs.~\onlinecite{Duchemin12,Palumno10}, intramolecular excitations, 
such as the $\pi$-$\pi^*$ Gouterman transitions in porphyrins, \cite{Gouterman} come well within 0.1 eV 
accuracy as compared to experiment, allowing to precisely predict the relative position of charge-transfer 
versus intramolecular excitations at a donor-acceptor interface.

\section{Results}

\begin{figure}
\begin{center}
\includegraphics*[width=0.45\textwidth]{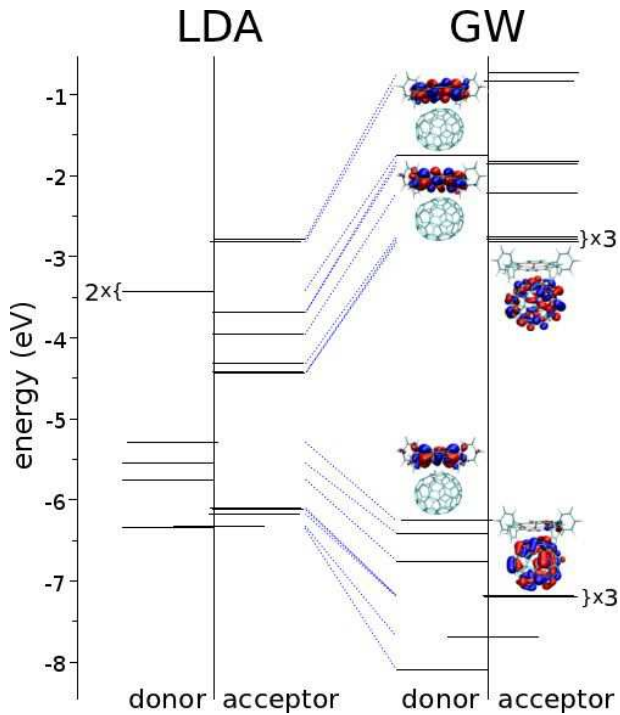}
\caption{ (Color online) Cartography of the Kohn-Sham (left) and $GW$ (right) eigenvalues. The energy 
levels are distributed into a (donor) or (acceptor) column according to the spatial localization of
the corresponding eigenstates. Frontier Kohn-Sham orbitals are represented next to their associated
energy level.}
\label{gw}
\end{center}
\end{figure}

We adopt the same geometry as that used in the constrained-DFT study of the $C_{70}$-ZnTPP complex
in Ref.~\onlinecite{Ghosh10}. Namely, the distance between the Zn atom and the center of the 
closest electron-rich (6:6)  $C_{70}$ carbon bond is fixed to 2.8~\AA\ using a constrained relaxation 
scheme. Such a geometry stems in particular from the analysis of the stability of several 
fullerene-porphyrin complexes in various geometries \cite{Wang02}. 
However, as shown recently in the case of a fullerenes-polythiophenes interface, \cite{McMahon11} 
significant disorder is to be expected at the bulk donor-acceptor interface, so that the present 
geometry is adopted for sake of comparison with available previous calculations of charge-transfer 
excitations. 

\subsection{Quasiparticle spectrum within $GW$}

We show in Fig.~\ref{gw} a cartography of the Kohn-Sham (left) and $GW$ (right) eigenvalues 
together with a representation of a few associated eigenstates.  As expected, the quasiparticle 
HOMO-LUMO gap is significantly opened from the DFT-LDA to the $GW$ calculation. The quasiparticle 
$GW$ HOMO-LUMO gap is found to be of 3.45 eV, namely 2.6 eV larger than the corresponding DFT-LDA 
Kohn-Sham result.  Our $GW$ HOMO-LUMO gap is in good agreement with the 3.56 eV value
found for the same system in Ref.~\onlinecite{Zope12} by taking differences of total energy 
between the neutral and positively or negatively charged complex at the DFT-PBE level within 
the standard $\Delta$SCF approach.  Similarly, our ionization potential of 6.25 eV is in close 
agreement with the reported 6.21 eV $\Delta$SCF-PBE value.  In the absence of gas phase experimental 
data, such an agreement between the two approaches is very satisfactory, even though $\Delta$SCF values 
are known to vary with the choice of the functional. \cite{GalliG0W0}

The donor-acceptor, or type-II, character of the complex is evidenced by the relative positions
(frontier orbitals energy offsets) of the $C_{70}$ and ZnTPP eigenstates across the gap. At the
$GW$ level, the ZnTPP HOMO (LUMO) is located 0.95 eV (1.05 eV) above the $C_{70}$ HOMO (LUMO) 
state. \cite{hybridization} The (HOMO) and (HOMO-1) orbitals are the standard porphyrin 
Gouterman orbitals \cite{Gouterman}, the (HOMO-2) being also localized onto ZnTPP, 
while the (nearly) 3-fold level beneath derive from the HOMO of the $C_{70}$ fullerene. 
\cite{Scuseria91,Saito91} The three lowest unoccupied levels, (LUMO) to (LUMO+2),
derive from the 2-fold $e_1$ and $a_1$ unoccupied levels of $C_{70}$, the two unoccupied
Gouterman orbitals located on ZnTPP being the (LUMO+6) and (LUMO+7) levels. Such a mapping
of orbitals will allow us to analyze the optical excitations in terms of intramolecular 
(Frenkel) versus charge-transfer transitions.

The intramolecular ZnTPP HOMO-(LUMO+6) gap is found to be 4.5 eV, of comparable size with the 
intramolecular C$_{70}$ (HOMO-3)-LUMO gap (4.4 eV), but $\sim$1.1 eV larger than the  
``charge-transfer" HOMO-LUMO quasiparticle gap. The relative strength of the intramolecular (Frenkel) 
and charge-transfer electron-hole pairs binding energy will reshuffle the ordering of the 
photo-induced transition as compared to the  $GW$ occupied to unoccupied quasiparticle energy 
differences. The comparison between the $\sim$ 3.5 eV HOMO-LUMO gap and the 1.6-1.8 eV optical absorption 
threshold found in Refs.~\onlinecite{Ghosh10,Mukherjee09} indicates that in such systems the electron-hole 
interaction is extremely large, amounting to about half the quasiparticle gap.  This is what we now study 
within the Bethe-Salpeter formalism that aims at accurately  describing the electron-hole interaction in 
the photo-induced excited states.

\subsection{The low-lying charge-transfer state within BSE}

\begin{figure}
\begin{center}
\includegraphics*[width=0.45\textwidth]{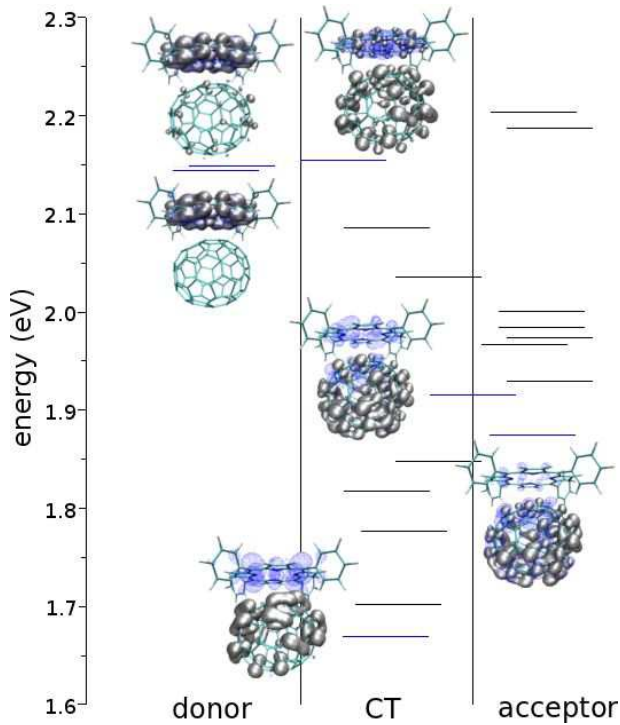}
\caption{ (Color online) Cartography of the BSE singlet neutral excitation energies. The transitions
are symbolically located from the left to the right in terms of their donor intramolecular, charge 
transfer (CT) or acceptor intramolecular character. In particular, transitions with a representative 
energy level displaced from the center to the left (right) indicates charge-transfer excitations with some 
weight onto intramolecular ZnTPP ($C_{70}$) transitions (see text). A few selected two-body eigenstates
are represented (see text and Fig.~\ref{bsestates}-caption for details).}
\label{bse}
\end{center}
\end{figure}

\begin{figure}
\begin{center}
\includegraphics*[width=0.45\textwidth]{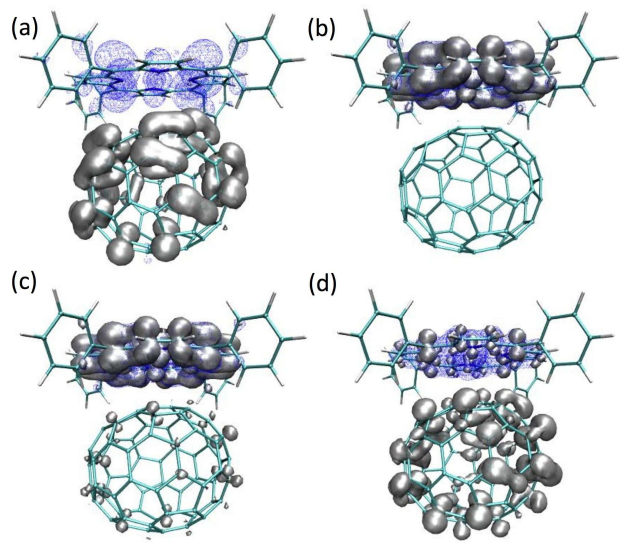}
\caption{ (Color online) Symbolic representation of the $\psi^{eh}_{\nu}({\bf r}_e,{\bf r}_h)$
two-body excitonic BSE eigenstates (see text), with (a) $\nu=0$ (CT$_0$), (b) $\nu=15$ 
(lowest intramolecular ZnTPP$^*$ excitation), and (c,d) $\nu=16$ and $\nu=17$ hot $CT_n$
excitations with both a charge-transfer and an intramolecular character.  The grey 
isocontours represent the hole-averaged electron distribution of charge and the wireframed 
blue contours the electron-averaged hole distribution. }
\label{bsestates}
\end{center}
\end{figure}

Our $GW$-BSE results for the neutral (singlet) excitation energies are provided in Fig.~\ref{bse} 
where we have distinguished the excitations with respect to their intramolecular or charge-transfer 
character. Such a distinction stems from the analysis of the weight of the Bethe-Salpeter two-body
eigenstates onto the Kohn-Sham occupied-unoccupied product state basis. We provide further, in 
Fig.~\ref{bse} and Fig.~\ref{bsestates}, a representation of selected two-body excitonic eigenstates 
$\psi^{eh}_{\nu}({\bf r}_e,{\bf r}_h)$, with $({\bf r}_e,{\bf r}_h)$ the electron and hole positions, 
respectively, and $\nu$ the excitation index, by plotting in grey the hole-averaged electron-distribution, 
that can be obtained by taking the expectation value of the electron $\delta({\bf r}-{\bf r}_e)$ position 
operator over $\psi^{eh}_{\nu}$. Similarly, the electron-averaged hole distribution is represented in 
blue (wireframe).

The $GW$-BSE low-lying singlet excitation is a clear charge-transfer state located at 1.67 eV
above the ground-state (we will use the notation $CT_0$ in the following).
 Since the C$_{70}$ LUMO is (nearly) 3-fold degenerate (see Fig.~1), 
the 3 low lying CT excitations are linear combinations of transitions between the HOMO and 
the LUMO, LUMO+1 and LUMO+2 levels, namely:

\begin{eqnarray*}
 {\psi}_{\nu=1,2,3}({\bf r}_e,{\bf r}_h)  & \simeq &  \sum_{n=0}^{2} c_n^{\nu}
   {\phi}_{homo}({\bf r}_h) \phi_{lumo+n}({\bf r}_e).
\end{eqnarray*}

\noindent For the lowest lying transition, we find typically: 
$(|c_1^1|^2=0.29, |c_2^1|^2=0.27, |c_3^1|^2=0.42)$. 
As such, these low-lying excitations are very clear charge-transfer states.  This reflects 
in the representation of the two-body wavefunction for the lowest $CT_0$ state 
(Fig.~\ref{bsestates}a) showing that the (hole-averaged) electron density is very clearly 
located onto the C$_{70}$ molecule while the (electron-averaged) hole density resides onto 
the donor ZnTPP molecule.

Our $\sim$1.67 eV $GW$-BSE excitation energy for the lowest excitation is in good agreement 
with the 1.8 eV value extrapolated from constrained-DFT calculation \cite{Ghosh10} under the
assumption that a full electron is transferred between the donor and the acceptor. On general
grounds, apart from the performances of the PBE functional in calculating differences of total
energies for such systems, a constrained approach within a variational framework necessarily 
yields an upper bound for the excitation energy.  As emphasized further in Ref.~\onlinecite{Ghosh10}, 
the assumption of a slightly reduced electron transfer would bring the constrained-DFT and BSE 
calculations in even better agreement. 
An analysis of Fig.~5 of Ref.~\onlinecite{Ghosh10} shows that the largest charge transfer 
explicitly studied by Ghosh and Gebauer corresponds to a transfer of $\sim$0.97 electron with 
a corresponding excitation energy of $\sim$1.7 eV, in close agreement with our findings, the
1.8 eV value being an extrapolated energy for full electron transfer. 

As observed in Refs.~\onlinecite{Ghosh10,Zope12}, the direct comparison with experiment is difficult.
Existing experimental values by Mukherjee and coworkers \cite{Mukherjee09} report the lowest excitation 
for the parent octadecyloxyphenyl-H$_2$TPP (freebase porphyrin) $C_{70}$ at 1.7 eV in toluene solvent 
and 1.8 eV in chlorophorme. As shown in Refs.~\onlinecite{Chukharev05,Zope12}, the replacement of the 
freebase porphyrin by a central Zn atom leads to a 0.1-0.2 eV redshift of the CT$_0$ excitation energy. 
On the contrary, removing the effect of solvent would lead to a compensating 0.2-0.3 eV blueshift (see 
analysis in Ref.~\onlinecite{Chukharev05}), even though the redshift of the CT excitation energy 
observed by Mukherjee and coworkers upon changing chlorophorm by the less polar toluene solvent 
indicates potentially more complex trends.  

Even though the comparison with experiment is difficult, we observe that the agreement is 
significantly better than what is obtained within TDDFT and (semi)local kernels,
namely $\sim$0.9 eV in the TD-PBE study of Ref.~\onlinecite{Ghosh10}.  The underestimation 
of charge-transfer excitations energy within TDDFT and (semi)local kernels is now well documented. 
Such a discrepancy lies in the lack of electron-hole interaction in the case of non-overlapping
HOMO-LUMO states, a situation that can be clearly evidenced e.g. by the absence of dispersion of the
CT excitation energy as a function of donor-acceptor distance, \cite{Dreuw03,Duchemin12} contradicting
the simple electrostatic limit of a (-1/D) scaling of the interaction between two well separated positive 
and negative charges, where D is some averaged measure of the electron-hole distance. The lack 
of electron-hole interaction for CT states indicates that the CT$_0$ excitation energy reduces to the
(too small)  Kohn-Sham HOMO-LUMO gap, that is indeed the $\sim$0.85 eV HOMO-LUMO gap that we find 
at the DFT-LDA level.

\subsection{Testing models for charge-transfer excitations}

The very clear charge-transfer character of the $CT_0$ low-lying excitation allows further to test 
common approximations by extracting from the BSE Hamiltonian the following matrix elements of the bare 
and (statically) screened Coulomb operator, namely:

\begin{eqnarray*}
 E^{B,(1)}_{eh} &=& \int \int  d{\bf r} d{\bf r}' |\phi_{homo}({\bf r})|^2  
      W({\bf r},{\bf r}') |\phi_{lumo+n}({\bf r}')|^2 ,  \\
 E^{B,(2)}_{eh} &=& \int \int  d{\bf r} d{\bf r}' |\phi_{homo}({\bf r})|^2  
      v({\bf r},{\bf r}') |\phi_{lumo+n}({\bf r}')|^2 , 
\end{eqnarray*}
 
\noindent
with (n=0,1,2). Such matrix elements are nearly independent of (n) with:
$E^{B,(1)}_{eh}$=1.56 $\pm$ 0.02 eV and $E^{B,(2)}_{eh}$=1.72 eV $\pm$ 0.06 eV. 
The exchange matrix elements, namely:

\begin{eqnarray*}
 E^{X}_{eh} &=& \int \int  d{\bf r} d{\bf r}' \phi_{h}({\bf r}) \phi_{l}({\bf r}) 
     v({\bf r},{\bf r}')  \phi_{h}({\bf r}') \phi_{l}({\bf r}') \\
\end{eqnarray*}

\noindent
where (h) stands for (homo) and (l) for either (lumo), (lumo+1) or (lumo+2), amount only to
less than 1 meV as a result of the charge separation between the (HOMO) and the (LUMO) states.
Subtracting now the screened Coulomb electron-hole interaction $E^{B,(1)}_{eh}$ to the
$GW$ quasiparticle HOMO-LUMO gap (3.45 eV), one obtains a simple estimation of the CT excitation
energy, namely 1.89 eV, in reasonable agreement 
with the 1.67 eV value obtained with the full $GW$-BSE treatment. As an even simpler approximation,
which does not necessitate the calculation of the full dielectric matrix, subtracting now the
bare Coulomb potential related $E^{B,(2)}_{eh}$ energy to the $GW$ HOMO-LUMO gap, one 
obtains 1.72 eV in (accidentally) even better agreement with the $GW$-BSE value. 

The present results clearly suggest that even in the limit of a separation of $\sim$2.8~\AA\
in a cofacial geometry, namely a distance and a configuration maximizing the wavefunction 
overlap between the donor and the acceptor for such $\pi$-stacking systems, the rather 
crude ``electrostatic" model 
consisting in replacing the electron-hole binding energy by the bare Coulomb integral 
$E^{B,(2)}_{eh}$ provides very acceptable results, with an error as compared to the 
full BSE value of the same order of magnitude than the difference between the $GW$-BSE 
treatment and the (extrapolated) constrained-DFT approach. This simple evaluation of the electron-hole 
binding energy in charge-transfer states was used for example in Ref.~\onlinecite{Isaacs11} 
to predict the evolution of the electron-hole binding energy for several donor-acceptor 
complexes and relative geometry. This simple scheme relies however on the assumption 
that the excited state is a clear charge-transfer state, with no mixing with higher 
energy $(\phi^e_i \phi^h_j)$ single-particle transitions, an approximation of the same 
nature than the one used in the constrained-DFT approach with the ``guess" that a full 
electron is transferred.

\subsection{``Hot" excited states: resonant intramolecular and charge-transfer excitations}

While our $GW$-BSE calculations yield results in good agreement with the computationally more 
efficient constrained-DFT approach for the low-lying $CT_0$ level, the $GW$-BSE framework provides 
the entire excitation spectrum as illustrated in Fig.~2. In particular, the relative position of 
the lowest intramolecular (Frenkel) donor excitation (labeled ZnTPP$^*$ here below), and the 
charge-transfer excitations at the donor-acceptor interface, is believed to be a crucial factor  
in understanding the mechanism(s) controlling the transition from a strongly bound electron-hole 
pair to a charge-separated (CS) state with unbound holes and electrons diffusing to the electrodes. 
In recent combined experimental and theoretical studies 
\cite{Muntwiler08,Ohkita08,Clarke08,Pensack09,Bakulin12,Murthy12} the role of hot CT$_n$ 
charge-transfer excitations in mediating the transition between Frenkel and CS states 
was advocated on the basis of experimental pump-probe experiments and quantum chemistry 
calculations.  Located at higher an energy than the lowest lying $CT_0$ level studied here 
above, the CT$_n$ hot states are significantly more delocalized due to their extra kinetic 
energy, a favorable feature for evolving into a charge-separated state.  

We can now analyze the present ZnTPP-C$_{70}$ system along these lines and study in particular the 
position of the ZnTPP$^*$ lowest intramolecular excitation with respect to the manifold of CT 
states. Within the present $GW$-BSE approach, the  ZnTPP$^*$ Q-like excited state \cite{Gouterman}
is located at $\sim$ 2.15 eV (see left ``donor" column in Fig.~\ref{bse}) with a corresponding 
two-body eigenstate provided in Fig.~\ref{bsestates}(b). Such a state is composed of transitions from 
the (HOMO-n) (n=0,1,2) ZnTPP-related levels to the (LUMO+6) and (LUMO+7) levels deriving from the lowest 
unoccupied levels of the isolated ZnTPP molecule (see Fig.~\ref{gw}). The $GW$-BSE excitation energy 
($\sim$  2.15 eV) is very close to the experimental $Q_{x,y}$ value of 2.09 eV for gas phase ZnTPP. 
\cite{Edwards77} Previous $GW$-BSE calculations of Q-transitions in a parent zincbacteriochlorin 
molecule \cite{Duchemin12} and freebase H$_2$TPP \cite{Palumno10} were shown to lead as well to an 
agreement within 0.1 eV as compared to experiment for such intramolecular low-lying excitations.

As depicted in Fig.~\ref{bse}, the ZnTPP$^*$  intramolecular excitation lies near or above several 
charge-transfer excitations located above the CT$_0$ state. Among these higher-energy, or hot, 
excited states, interesting transitions are the excitations $(\nu=16)$ and 
$(\nu=17)$ that lies within 10 meV of the  ZnTPP$^*$ lowest intramolecular excitation. With the
notations defined above, one finds for the corresponding two-body eigenstates:

\begin{eqnarray*}
{\psi}_{\nu=16,17}({\bf r}_e,{\bf r}_h) 
       &  \simeq & c_{h,l+6}^{\nu} \; {\phi}_{homo}({\bf r}_h)  {\phi}_{lumo+6}({\bf r}_e)  \\
       &   +    &  c_{h-1,l+7}^{\nu} \;  {\phi}_{homo-1}({\bf r}_h)  {\phi}_{lumo+7}({\bf r}_e) \\
       &   +    &  c_{h-2,l}^{\nu} \; {\phi}_{homo-2}({\bf r}_h)  {\phi}_{lumo}({\bf r}_e),
\end{eqnarray*}

\noindent with $( |c_{h,l+6}^{\nu}|^2, |c_{h-1,l+7}^{\nu}|^2, |c_{h-2,l}^{\nu}|^2 )$ equal
to ($0.40,0.32,0.15$) for ($\nu=16$) and ($0.10,0.09,0.74$) for ($\nu=17$). Again, 
the (lumo+6) and (lumo+7) levels are mainly the nearly degenerate lowest unoccupied
Gouterman orbitals of the ZnTPP molecule. As such, the two-body wavefunctions present both a charge-transfer
character, with the (homo-2) to (lumo) transition component, but also an intramolecular ZnTPP character, with
the (homo) to (lumo+6) and (homo-1) to (lumo+7) components.  The corresponding excitonic wavefunctions 
are represented in Fig.~3(c,d), evidencing this hybrid character.  The $(\nu=16)$ excitation 
is mainly a ZnTPP intramolecular transition showing however a significant probability for the electron
(grey contours) to jump onto the $C_{70}$ acceptor. The $(\nu=17)$ excitation is mainly a CT transition
with the excited electron sitting mainly onto the $C_{70}$ acceptor, but with a strong 
weight as well onto the ZnTPP donor, in great contrast with the low lying CT$_0$ state which is a pure 
charge-transfer state. 

While the present system, comprising only one donor-acceptor complex, cannot be used to test the 
delocalization of the hot excited states over several donor (for the hole) or acceptor (for the 
electron) molecules, a delocalization expected to favor charge separation,  
the present finding indicates that the extra kinetic energy of the hot excited states, as compared 
to the low lying CT$_0$ state, induces also the delocalization of the hole and electron over the 
neighboring donor and acceptor molecules, generating hybrid intramolecular and charge-transfer 
states. The present results are consistent with a previous semi-empirical  study \cite{Aryapour10} 
of a $C_{60}$-polymer blend evidencing a rapid drop of the ``ionicity" of charge-transfer excited states 
with increasing energy, a reduction of the charge-transfer character characterizing an hybridization 
with intramolecular transitions. It was shown however that the number, energy and character of the
CT states lying below the intramolecular donor (polymer) excitations were strongly dependent on the 
parametrization of the intersite ($V_{ij}$) Coulomb potential, making the case for accurate 
parameter-free \textit{ab initio} calculations.

\begin{figure}
\begin{center}
\includegraphics*[width=0.45\textwidth]{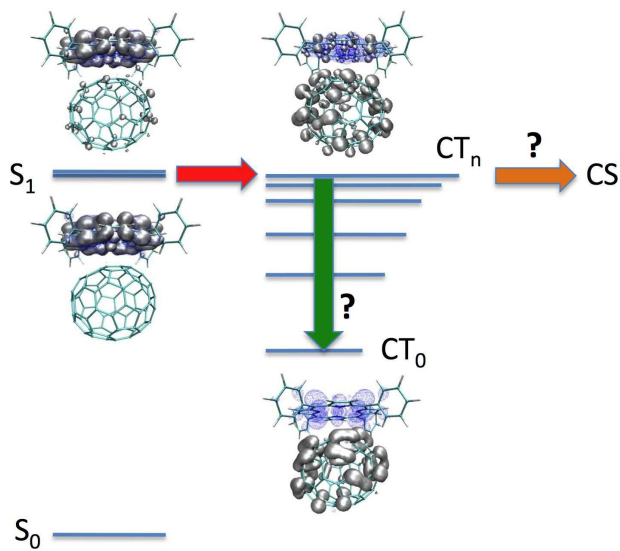}
\caption{(Color online) Symbolic representation (following Ref.~\onlinecite{Bakulin12}) showing 
the potential flow of transitions from intramolecular (Frenkel) excitations, localized on the 
donor, to resonant hot charge-transfer (CT$_n$) states (red arrow) and subsequently unbound 
charge-separated (CS) electron-hole pair (orange arrow). The adverse decay into the low lying 
CT$_0$ state is also represented with the green arrow.}
\label{cartoon}
\end{center}
\end{figure}

Such an energy alignment and wavefunction overlap between the ZnTPP$^*$ intramolecular low-lying excitation 
and these hot CT$_n$ states is a strong indication that transitions between these resonating states will 
be strongly favored. \cite{Kawatsu08} The present findings may suggest an additional criteria for 
designing donor-acceptor complexes with efficient charge separation through transient hot (delocalized
and weakly bound) charge-transfer states,  namely the existence, at the energy of the lowest ZnTPP$^*$
donor intramolecular excitation, of a large density of resonating states with hybrid intramolecular and 
charge-transfer character. Adopting a simple hydrogenoid model for the electron-hole pair, this can be 
expected to be achieved by maximizing the energy difference between the low-lying $CT_0$ charge-transfer 
state and the intramolecular ZnTPP$^*$ donor excitation. Even though equivalent to the standard criteria
for energy-gradient-driven transitions to the CT$_0$ state, we emphasize that the argument here applies 
to a scenario where electron-hole dissociation occurs through hot charge-transfer states, avoiding 
possibly the adverse thermalization into the low lying strongly bound CT$_0$ exciplex.  
\cite{Bakulin12,Muntwiler08,Yi09} This observation, together with the variety of systems, and 
possible differences in cristallinity, on which pump-probe experiments have been performed, 
certainly explain that the role of hot versus cold charge-transfer states in mediating 
electron-hole separation remains controversial.
Following Ref.~\onlinecite{Bakulin12}, we represent schematically in Fig.~\ref{cartoon} a flow of 
transitions from donor intramolecular to resonant hot CT$_n$ states, followed either by the relaxation
towards the low-lying CT$_0$ state or the direct transition to a hot charge-separated (CS) state.

In a recent $GW$-BSE study of a $C_{60}$-quaterthiophene system, \cite{Baumeier12} much 
emphasis was given to calculating the difference of energy (${\Delta}{\Omega}$) between the lowest 
donor intramolecular Frenkel (FE) exciton and the charge-transfer CT$_0$ state. ${\Delta}{\Omega}$ 
was found to be sensitive to the relative donor-acceptor orientation, with an average positive 
value strongly detrimental to the $FE \rightarrow CT_0$ transition. Further, as observed here above, 
the binding energy ($E_B^{CT_0}$) of the electron-hole pair in the $CT_0$ state was found to be 
much larger than room temperature. Even though based on very accurate calculations, such results 
do not allow to understand the observed efficiency of state-of-the-art fullerene-polythiophene 
systems.  While the difference between isolated gas phase donor-acceptor dimers, on which 
calculations are usually performed, and more realistic bulk donor-acceptor heterojunctions 
(BHJ) was invoked, \cite{Baumeier12} a certainly crucial issue, a possible alternative interpretation 
lies in the scenario where charge separation occurs directly through hot CT$_n$ states, without 
transiting to the highly bound CT$_0$ state, so that variations in  ${\Delta}{\Omega}$ and 
$E^B_{CT0}$ become less relevant.  

Back to the present ZnTPP-C$_{70}$ system, the further analysis of the CT states lying between 
the CT$_0$ and ZnTPP$^*$ levels shows several states with hybrid charge-transfer and intramolecular 
character, but with a clear weight onto intra-fullerene excitations. Since photoexcited pairs are 
mainly generated in the donor (ZnTPP) part of the heterojunction, such excitations cannot be efficient 
intermediate states for charge separation, indicating on the contrary a pathway to an adverse 
relocalization of the electron-hole pair onto the fullerene side.

As a final analysis, we now compare our results to a recent study of the excited states of fullerene-porphyrin
complexes \cite{Zope12} based on another computationally efficient DFT-based framework. As described in 
Ref.~\onlinecite{Baruah09}, this efficient, but approximate, approach relies on a second-order perturbative 
correction to the ``excited" Slater determinant obtained by replacing one occupied single-particle DFT 
Kohn-Sham orbital by an empty one in the ground-state Kohn-Sham many-body wavefunction. Since the empty and
occupied orbitals considered were the (HOMO) to (HOMO-2) and (LUMO) to (LUMO+2) orbitals, only low-lying 
charge-transfer states could be studied, excluding in particular the intramolecular ZnTPP excitations, 
involving the (LUMO+6) and (LUMO+7) single-particle empty orbitals, and the hybrid states above-mentioned.

The lowest charge-transfer $CT_0$ excitation within this scheme was found to be located at 1.95 eV above 
the ground-state.  \cite{Zope12} Our $GW$-BSE value is $\sim$ 0.3 eV smaller.  Even though slightly less 
satisfactory than the comparison with the variational constrained DFT calculation \cite{Ghosh10}, the 
agreement remains reasonable. Since this approach is perturbative, it is difficult to conclude whether, 
on general grounds, it should produce an upper- or lower-bound for the exact result. In however excellent 
agreement with this perturbative technique, the three lowest excitations are CT excitations with a strong 
weight on the HOMO to LUMO, LUMO+1 and LUMO+2 quasidegenerate levels, the following excitation being mainly 
a (HOMO-1) to (LUMO+2) transition. Inclusion of intramolecular transitions in the pool of studied excitations 
may certainly help in favoring the comparison with other techniques and in revealing the excitations with 
hybrid intramolecular and charge-transfer character.

\section{Conclusion}

We have studied within the many-body Green's function $GW$ and Bethe-Salpeter formalisms the singlet
excitation energies  in a fullerene-porphyrin $C_{70}$-ZnTPP complex. The lowest-lying excited state is 
a charge-transfer (CT$_0$) state with a $\sim$ 1.67 eV energy, a value in good agreement with recent 
constrained DFT calculations leading, for the same geometry, to a 1.8 eV upperbound in the asymptotic
limit of a full electron transfer. The energy of the CT$_0$ state can be estimated simply from the
knowledge of the correct $GW$ HOMO-LUMO gap and of simple Coulomb integrals between the HOMO and LUMO
Kohn-Sham eigenstates density.  Beyond the lowest excitation, the $GW$-BSE framework provides the full 
excitation spectrum, allowing in particular to assess the relative position of the lowest intramolecular
ZnTPP$^*$ donor excitation with respect to the manifold of hot charge-transfer $CT_n$ states.
We reveal in particular
the existence of several excitons with an hybrid intramolecular and charge-transfer character which are
resonant in energy with the ZnTPP$^*$ excitation. Such findings suggest a rapid transition from the
photo-generated ZnTPP$^*$  state to hot CT$_n$ states with excess kinetic energy as compared to the 
lowest CT$_0$ state, favoring possibly the transition towards unbound charge-separated states. More direct
comparison to experimental results requires to account for environmental effects, namely the renormalization
of the quasiparticle and neutral excitation energies from the gas phase to realistic bulk phases. Further, 
a conclusive study of the evolution of intramolecular exciton into delocalized ``band" excitation certainly
require the ability to treat extended systems, beyond  cluster calculations. The combination of the present 
$GW$ and Bethe-Salpeter formalisms with proper and accurate polarizable models is certainly
an important and ambitious goal for the community in order to properly discriminate between the various 
scenario proposed for charge separation in a realistic environment. 

\appendix*
\section{Convergence tests}

We provide here below in Table I a comparison of the results obtained using a smaller Kohn-Sham basis,
namely a triple-zeta plus single polarization (TZP) basis, instead of the TZ2P basis used here above.
We focus on the lowest lying CT$_0$ charge-transfer state,
the two lowest lying nearly degenerate intramolecular ZnTPP$^*$ states, and the resonant CT$_n$ state,
discussed in the text. Differences between the two calculations are seen to be smaller than 15 meV,
indicating the quality of the convergence for the present $GW$-BSE level at the chosen TZ2P level.
We further test the number of empty states used at the Bethe-Salpeter level (the full unoccupied spectrum 
is used at the $GW$ level). Clearly, as expected, the charge transfer states converge very quickly due to their
large weight onto states around the energy gap. The Frenkel excitations, with more mixing with higher states,
show a small 20 meV shift upon increasing the number of empty states from 240  to the 360 states used 
in the present study. 

\begin{table}
\begin{tabular}{l|c|cccc}
\hline
             &     &   CT$_0$    &  ZnTPP$^*_1$ & ZnTPP$^*_2$ &  CT$_n$  \\
              &    &   ($\nu$=1) &  ($\nu$=15)  &  ($\nu$=16) & ($\nu$=17) \\
\hline
   TZP     &    240   &   1.678  &  2.179       &   2.182    &  2.174    \\
   TZP     &    360   &   1.677  &  2.160       &   2.163    &  2.176    \\
\hline
   TZ2P   &    360        &   1.670  &  2.147       &   2.149    &  2.155    \\
\hline
\end{tabular}
\caption{ Excitation energies for selected states comparing TZP and TZ2P bases. 
The number of empty states for the BSE Hamiltonian is further indicated (second column).
Energies are in eV. The excitation ($\nu$) index is indicated (see text).  }
\label{table}
\end{table}

\textbf{Acknowledgements} 
The authors are indebted to R.~Gebauer for providing the geometry of the ZnTPP-C$_{70}$ complex studied
here above.  Calculations have been performed on the CEA-Curie supercomputing facility thanks to 
national GENCI and european PRACE (no. 2012071258) projects.
I.D. acknowledges partial funding from the CEA Eurotalent program and X.B. from the French Research Agency 
under contract ANR 2012-BS04 PANELS.  The authors acknowledge C.~Attaccalite, V.~Olevano and C.~Faber for 
a critical reading of the manuscript.



\begin{thebibliography}{150bliography}
\bibitem{basis2}
Concerning the specificities of the DFT {\sc{Siesta}} package used to generate
the input Kohn-Sham states, the orbital confinement constraint was  released by 
imposing an arbitrary large confinement radius and the generated numerical atomic
orbitals were fitted by up to 8 (contracted) Gaussians.
\end{thebibliography}
\end{document}